\documentclass[pra,showpacs,twocolumn,superscriptaddress,longbibliography]{revtex4-2}
\usepackage{amsmath}   
\usepackage{amssymb}
\usepackage{graphicx}
\usepackage{dcolumn}
\usepackage{bm}
\usepackage{amsmath}
\usepackage{graphicx}
\usepackage{amsfonts}
\usepackage{subfigure}
\usepackage{graphicx}
\usepackage{float}
\usepackage{color}
\usepackage{xcolor}
\usepackage{soul}
\usepackage[normalem]{ulem}
\usepackage{braket}
\usepackage{amssymb}
\usepackage{hyperref}
\usepackage{comment}
\usepackage{xr}
\usepackage[T1]{fontenc}

\hypersetup{
    colorlinks=true,
    citecolor=black,
    linkcolor=black,
    filecolor=black,      
    urlcolor=black,
}
\begin{document}
\title{Coupling single atoms to a nanophotonic whispering-gallery-mode resonator via optical guiding}
\author{Xinchao Zhou}
\affiliation{Department of Physics and Astronomy, Purdue University, West Lafayette, IN 47907, USA}
\author{Hikaru Tamura}
\affiliation{Department of Physics and Astronomy, Purdue University, West Lafayette, IN 47907, USA}
\author{Tzu-Han Chang}
\affiliation{Department of Physics and Astronomy, Purdue University, West Lafayette, IN 47907, USA}
\author{Chen-Lung Hung}
\email{clhung@purdue.edu}
\affiliation{Department of Physics and Astronomy, Purdue University, West Lafayette, IN 47907, USA}
\affiliation{Purdue Quantum Science and Engineering Institute, Purdue University, West Lafayette, IN 47907, USA}
\date{\today }

\begin{abstract} 
We demonstrate an efficient optical guiding technique for coupling cold atoms in the near field of a planar nanophotonic circuit, and realize large atom-photon coupling to a whispering-gallery mode in a microring resonator with a single-atom cooperativity $C\gtrsim 8$. The guiding potential is created by diffracted light on a nanophotonic waveguide that smoothly connects to a dipole trap in the far field for atom guiding with subwavelength precision. 
We observe atom-induced transparency for light coupled to a microring, characterize the atom-photon coupling rate, extract guided atom flux, and demonstrate on-chip photon routing by single atoms. Our demonstration promises new applications with cold atoms on a nanophotonic circuit for chiral quantum optics and quantum technologies.
\end{abstract}

\maketitle

Ultracold atoms strongly coupled to photonic fields are model systems for realizing quantum nonlinear optics \cite{2014_NaturePhotonics_Chang}, quantum networks \cite{2008Nature_QuantumInternet_Kimble, RMP2015}, and quantum simulations of many-body physics \cite{RMP2013,2015NaturePhotonics_manybody,2015NaturePhotonics_subwavelength}.
Interfacing cold atoms with nanoscale photonic waveguides \cite{Hinds2011, twocolor2010,twocolor2012,2016PRL_BraggScattering_Appel,2016PRL_BraggScattering_Laurat,2017NC_Superradiance_Rolston,2019Nature_Laurat_atomarray,2014NC_Goban,PRL2015_Superradiance} and resonators \cite{2013Nature_Lukin,2013PRL_Rauschenbeutel,2014Nature_Lukin,2014NaturePhotonics_phasegate_Rauschenbeutel,2015PRL_Aoki,2016Science_Circulator,2018NaturePhysic_Dayan,2019PRL_nanoFiberCavity_Nayak,2019PRL_aoki,2020PRL_Lukin,2020PRL_Rauschenbeutel_WGM,2021Science_Lukin} in quasi-linear (1D) and planar (2D) geometries further promises stronger atom-light interactions and novel quantum functionalities via dispersion engineering, controlled photon propagation, topology, and chiral quantum transport, thus leading to new paradigms for quantum optics beyond conventional settings in cavity and waveguide quantum electrodynamics (QED) \cite{RMP2018,2021review_waveguideQED}. 

To date, the key challenge for atom-nanophotonic integration remains to be efficient transporting and trapping of cold atoms on nanoscale dielectrics. 
Success so far has been limited to suspended 1D structures, which are surrounded by vacuum and allow for laser-cooled atoms to be loaded directly into optical traps in the near field (distance $z\lesssim$ optical wavelength above surface). Examples include optical nanofibers \cite{twocolor2010,twocolor2012}, where an array of atoms can be localized in a lattice of two-color evanescent field traps formed by guided light. Through external-illumination, a tight optical trap can also form on top of a suspended waveguide \cite{2013Nature_Lukin,PRL2015_Superradiance,2019PRL_nanoFiberCavity_Nayak}. For deterministic atom trapping, optical tweezers or an optical conveyor belt have been utilized to initiate atom loading in freespace, followed by transport to a proximal photonic crystal \cite{2013Nature_Lukin, 2019PNAS_Clocked_Delivery_Kimble}. These guiding and trapping techniques enable demonstrations of cooperative atom-photon coupling \cite{PRL2015_Superradiance, 2016PRL_BraggScattering_Appel,2016PRL_BraggScattering_Laurat,2017NC_Superradiance_Rolston}, and collective Lamb shifts with trapped atoms \cite{2016PNAS_Atom-atom_Hood}. Waveguide-interfaced atomic quantum memories \cite{2019Nature_Laurat_atomarray}, photonic phase gate \cite{2014Nature_Lukin}, and atom-photon/atom-atom entanglement \cite{2021Science_Lukin} have also been realized. 

Extending optical trapping to 2D photonic structures, however, faces immediate challenges. Due to restricted trap opening to freespace and reduced laser cooling efficiency in the near field above a dielectric plane, potentially caused by unbalanced radiation pressure from surface reflection and scattering or from increased heating rates due to mechanical vibrations \cite{2019PRXHeating}, unobstructed atom loading into a near field trap has shown limited success probability \cite{2019NC_MayKim, 2021SPIE_Hikaru}. This has prevented further explorations of atom-light coupling on more complex and interesting planar structures such as 2D photonic crystals \cite{2019PNAS_2DPhC} and whispering-gallery mode (WGM) microring resonators with propagation-direction-dependent, chiral atom-light interactions \cite{2017Nature_Chiral,2013PRL_Rauschenbeutel,2014Science_Dayan}. Without tackling the challenges of cooling and trapping, thermal atomic vapors have already been coupled to integrated ring resonators \cite{2016Levy_thermalatoms, 2016NJP_thermalatoms} and waveguides \cite{2007NP_Hawkins, 2017NC_atomiccladdingwaveguides, 2018PRX_thermalatoms_slotwaveguides, 2022PRR_thermalatoms_slotwaveguides}, but with much limited single-atom interaction time and cooperativity.
 
\begin{figure*}[!t]
\centering
\includegraphics[width=0.99\textwidth]{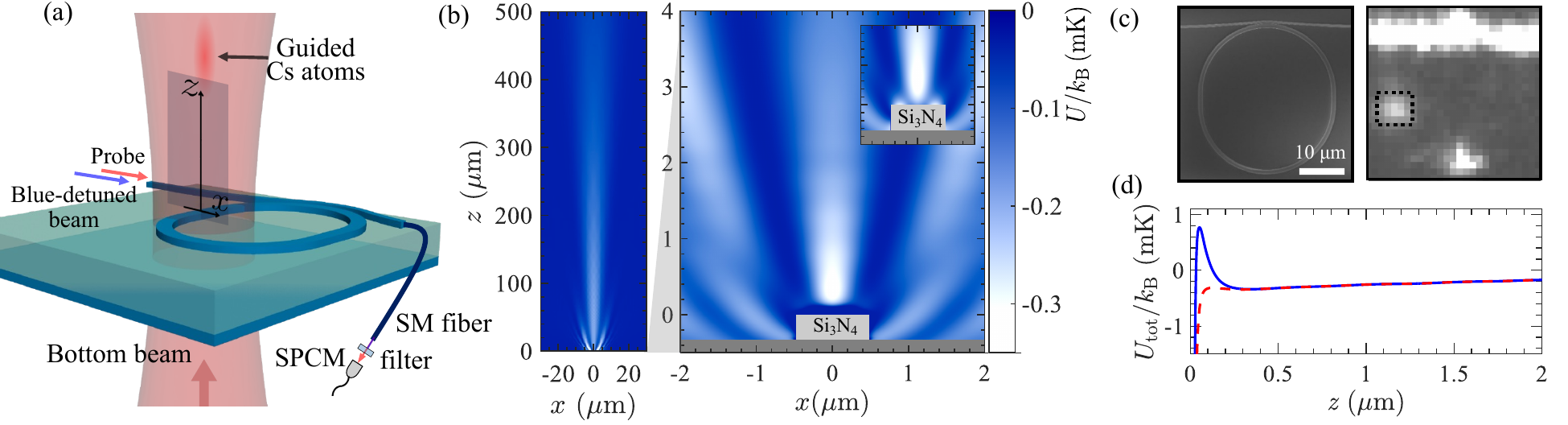}
\caption{Optical funnel on a nanophotonic microring circuit.
(a) Schematic of the setup. An optical funnel is formed by a red-detuned, bottom-illuminating beam. 
WGMs in the microring are excited by a probe field co-propagating with a blue-detuned beam.
The latter forms a repulsive potential barrier to plug the optical funnel. Transmitted light is directed to a single photon counting module (SPCM) after wavelength filtering. (b) Cross section of the funnel potential $U(x,z)$ (left) and a zoom in view near the Si$_3$N$_4$ waveguide (right). 
Inset shows the potential without a barrier (unplugged funnel). Gravity is along the $-z$ direction.
(c) Scanning electron micrograph of a microring (left), and an optical micrograph (right, same field of view) showing fluorescence from guided atoms (bounded by dashed box). Other bright spots are unfiltered scattered light from the waveguide. (d) Potential line cuts $U_\mathrm{tot}(0,z)$ with (solid curve) and without (dashed curve) the repulsive barrier. $U_\mathrm{tot}=U+U_\mathrm{cp}$ includes the atom-surface Casimir-Polder potential $U_\mathrm{cp}$ \cite{SM}.}
\label{fig:fig1}
\end{figure*}

In this letter, we overcome such restrictions using a technique for precision guiding of cold atoms from far field ($z \gtrsim 250\,\mu$m) to a nanoscale optical trap in the near field with subwavelength precision. This scheme is projected to work with generic dielectric nanostructures -- a far-off resonant optical beam forms a tapered guiding potential towards a bottom-illuminated structure (Fig.~\ref{fig:fig1}), where diffracted light in the near field can precisely direct trapped atoms towards the surface like a geometrically defined `optical funnel'. We show that the end of an optical funnel ($z\lesssim 100\,$nm) can be plugged using a repulsive evanescent field potential that could also counteract atom-surface Casimir-Polder attraction to form a stable trap [Fig.~\ref{fig:fig1}(d)].
We implement an optical funnel for guiding cold atoms and coupling them, for the first time, to a nanophotonic microring resonator in a fiber-integrated circuit \cite{3Dtaper,2020APL_Chang,SM}. We achieve synchronous atom guiding towards a designated spot on a microring, and report observation of atom-photon coupling in a WGM of the microring resonator with single-atom cooperativity $C\gtrsim 8$ during the atom-transit. We extract a peak atom flux $\approx 240\,$ms$^{-1}$, under a peak atom-photon coupling rate $g_\mathrm{max}/(2\pi) \approx 136\,$MHz for a spin-polarized atom. 
Our scheme is complementary to an optical conveyor belt \cite{2019NC_MayKim, 2019PNAS_Clocked_Delivery_Kimble}, and can be extended to guiding and trapping atom arrays in generic planar nanostructures. 

We begin the experiment by collecting $\sim 10^5$ laser-cooled cesium atoms (temperature $\sim 20\,\mu$K following polarization gradient-cooling, PGC) at $z\approx 250\,\mathrm{\mu m}$ above a transparent silica membrane that hosts a racetrack-shaped Si$_3$N$_4$ microring resonator \cite{2019Optica_Chang, SM}. The atoms are loaded into an optical funnel that points towards the microring waveguide of width $\approx 950\,\mathrm{nm}$ and height $\approx 326\,\mathrm{nm}$, respectively. The funnel potential is formed in a red-detuned, bottom-illuminating beam (wavelength $\lambda_\mathrm{r} \approx 935.3\,\mathrm{nm}$), with a beam waist of $7\,\mathrm{\mu m}$ and a polarization locally parallel to the waveguide. Over the top, the zeroth-order diffraction exhibits strong intensity gradient, diffracting from a 200\,nm $1/e^2$-transverse width into a circular far-field dipole beam profile; see Fig.~\ref{fig:fig1}(b) and \cite{SM}. This guiding potential in the near field is robust against beam misalignment by more than the width of the microring waveguide ($\gtrsim 1~\mu$m), which we confirmed in simulation and experimentally. Higher order diffractions do not form funnels because they display intensity maxima that are several micrometers away from the waveguide. Localized atoms in the optical funnel can be fluorescence imaged at distances $z \lesssim 10\,\mu$m \cite{2019NC_MayKim,2021SPIE_Hikaru}, as shown in Fig.~\ref{fig:fig1}(c).

We plug the optical funnel using a repulsive evanescent field formed by a `blue' WGM 
(wavelength $\lambda_\mathrm{b} \approx 849.55\,\mathrm{nm}$). A plugged funnel potential exhibits a stable trap minimum in the near field. We adjust the power of the bottom beam, $P_\mathrm{r}\approx 15\,\mathrm{mW}$, and that of the blue-detuned beam, $P_\mathrm{b}\approx 33\,\mu\mathrm{W}$, to form a closed trap with trap minimum at $z\approx 280\,$nm and a trap depth of $k_{\mathrm{B}}\times 250\,\mu$K, where $k_{\mathrm{B}}$ is the Boltzmann constant; see Fig.~\ref{fig:fig1}(d).

We detect guided atoms in the near field by probing atom-WGM photon interactions. The `probe' WGM resonance is thermally stabilized to the $F=4 \leftrightarrow F'=5$ transition in D2 line. Probe photons are sent through one end of a bus waveguide to couple to the clockwise circulating (CW) WGM (coupling rate $\kappa_{\mathrm{e}}\approx 2\pi \times 0.77\,$GHz). The intrinsic photon loss rate is $\kappa_{\mathrm{i}}\approx 2\pi \times 0.95\,$GHz, and the total photon loss rate is $\kappa=\kappa_{\mathrm{e}}+\kappa_{\mathrm{i}}\approx 2\pi \times 1.72\,$GHz ~\cite{SM}. Resonant probe photons are drawn into the microring and dissipate, reducing the bus waveguide transmission to $T_{0}=|(\kappa_{\mathrm{e}}-\kappa_{\mathrm{i}})/\kappa|^2\approx 0.01$. Interaction with an atom will lead to an increased transparency $T>T_0$ \cite{2006Nature_Kimble, SM}. We note that a WGM photon is nearly circular-polarized in the near field. 
Interaction with the probe WGM (in CW circulation) can thus drive $\sigma^{+}$ transitions with spin axis defined transversely to the waveguide; an atom can also emit a photon in the counter-clockwise circulating (CCW) WGM via the $\sigma^{-}$ transitions, inducing reflection in the bus waveguide; see discussions in \cite{SM}. In our microring, CW and CCW WGM resonances are degenerate.

\begin{figure}[!t]
\centering
\includegraphics[width=1\columnwidth]{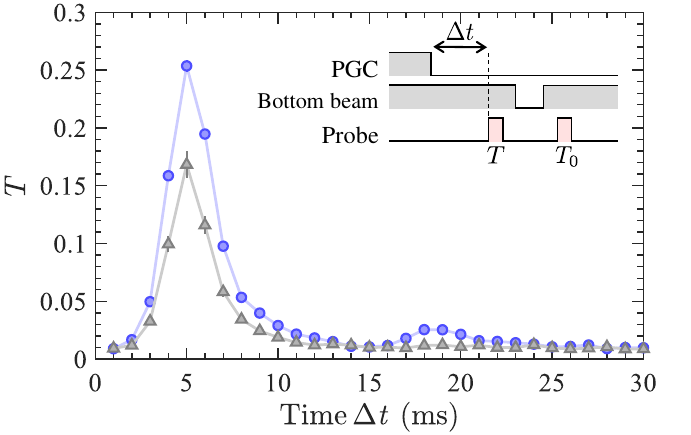}
\caption{Atom guiding in the optical funnel. Resonant transmission $T$ versus guiding time $\Delta t$ in a plugged (circles) and unplugged (triangles) funnel, respectively. Experimental sequence is illustrated in the inset. }
\label{fig:fig2}
\end{figure}

Our probe sequence is illustrated in Fig.~\ref{fig:fig2} inset. We prepare atoms in the $F=4$ ground state and then shut off the PGC light, allowing them to be guided towards the microring surface. After a wait time $\Delta t$, two weak probe pulses are sent through the bus waveguide, each with a duration of $1\,\mathrm{ms}$, to measure the transmission $T$ ($T_0$) in the presence (absence) of atoms; the bottom beam is switched off for $3\,\rm{ms}$ between the two pulses to release guided atoms. Each experiment is repeated $100$ times for averaging.

In Fig.~\ref{fig:fig2}, we observe increased transmission and a clear maximum up to $T \approx 0.26$ at $\Delta t \approx 5\,\mathrm{ms}$, indicating a peak atom flux arriving at the near field. Interestingly, transmission resurges at $16\,{\mathrm{ms}} \lesssim \Delta t\lesssim 21 \,$ms. This is due to longitudinal reflection of most guided atoms in a rapidly narrowing optical funnel \cite{2002PRA_adiabatic_theorem,2011PRA_funnel}, regardless of the presence of the repulsive potential barrier. These reflected atoms are later drawn back towards the surface for recoupling. We have performed atomic trajectory simulations \cite{note_sim} to confirm the observed oscillatory behavior and guiding effect \cite{SM}.

\begin{figure}[!t]
\centering
\includegraphics[width=1.0\columnwidth]{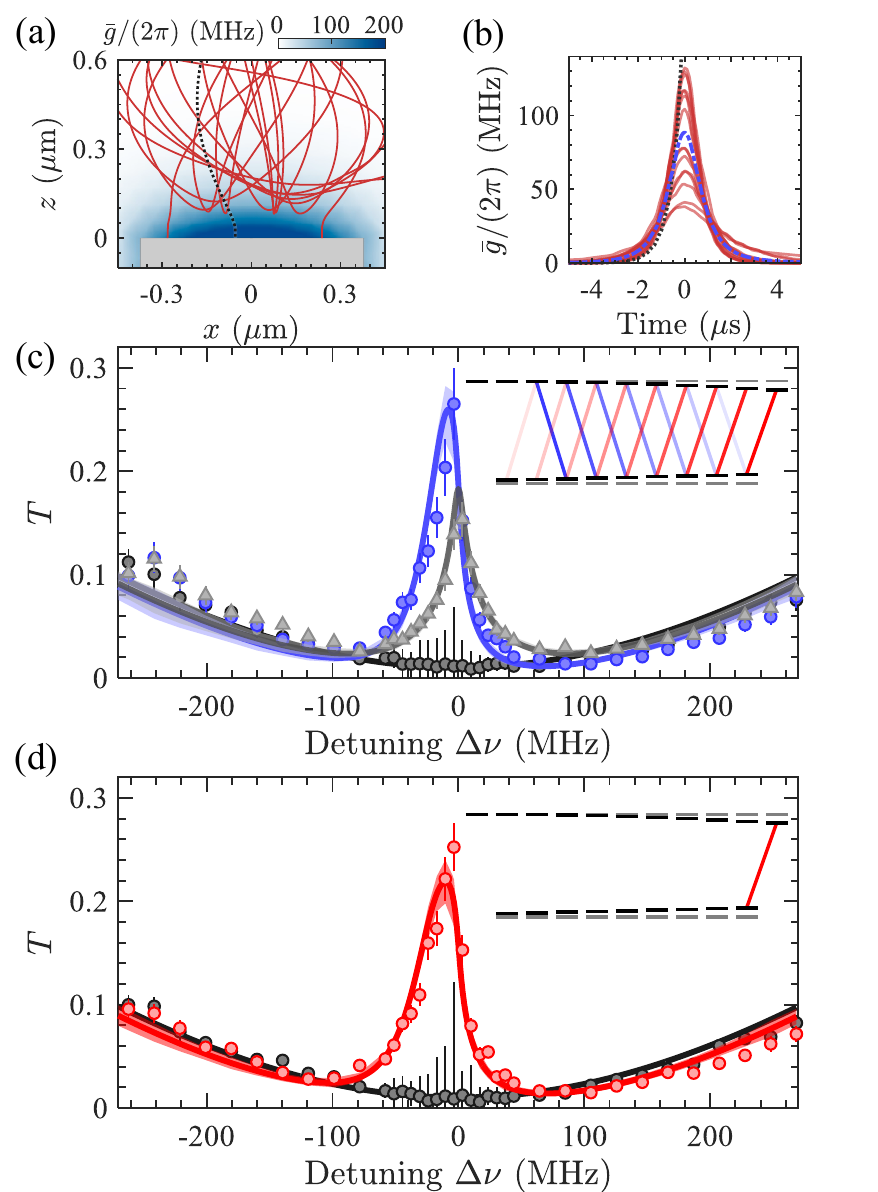}
\caption{Atom-WGM photon interaction. (a) Sample trajectories (solid curves) in the near field. False color map shows unpolarized atom-photon coupling strength $\bar{g}$. (b) $\bar{g}$ versus time for sample trajectories; blue dash-dotted curve is the mean. Black dotted curves in (a-b) show a typical case without the repulsive barrier. The time origin is aligned with the time to have the largest coupling strength for each trajectory. (c) Measured transmission $T$ versus laser detuning $\Delta \nu$ for unpolarized atoms with (blue circles) and without (gray triangles) the repulsive barrier, and $T_{0}$ for bare resonator without atoms. Solid blue (gray) curve is a single parameter fit using Eq.~(\ref{eq_t}) and input from trajectory calculations as in (a-b) with (without) the repulsive barrier. 
Shaded band shows $95\%$ pointwise confidence level. (d) Measured and fitted transmission $T$ versus laser detuning $\Delta \nu$ for polarized atoms in a plugged funnel. Insets in (c-d) illustrate the levels involved in the $F=4\leftrightarrow F'=5$ transition.}
\label{fig:fig3}
\end{figure}

We note that transparency is more pronounced with coupling to guided atoms in a plugged optical funnel, due to the fact that a repulsive barrier can increase the atom-WGM photon interaction time. To see this, in Fig.~\ref{fig:fig3}(a) we overlay sample atomic trajectories and position-dependent atom-photon coupling strength $\bar{g}$, calculated using the CW WGM field distribution \cite{comsol} and averaged over $g$ of all magnetic sub-levels \cite{SM}. Note that $\bar{g}$ is constant along the waveguide ($y$-)axis. Most trajectories exhibit a longitudinal classical turning point in the near field $z=110\pm 20\,$nm and within $|x|\lesssim 0.3\,\mathrm{\mu m}$. 
Corresponding time-dependent coupling strengths $\bar{g}(t)$ are plotted in Fig.~\ref{fig:fig3}(b). Averaging over all trajectories, we find peak $\bar{g}_{\mathrm{max}} \approx 2\pi \times 97\,\mathrm{MHz}$, corresponding to a peak cooperativity $\bar{C}^+ = 4\bar{g}_{\mathrm{max}}^2/\kappa/\gamma\approx 4.2$. Here $\gamma=2\pi\times5.2~$MHz is the atomic decay rate in freespace. The averaged interaction time (root-mean-square time weighted by interaction strength) would approach $t_i^\mathrm{b} \approx 2\,\mathrm{\mu s}$, more than doubled from $t_{i} \approx 0.9\,\mathrm{\mu s}$ for typical trajectories without a repulsive barrier. 

\begin{figure*}[!t]
\centering
\includegraphics[width=1.0\textwidth]{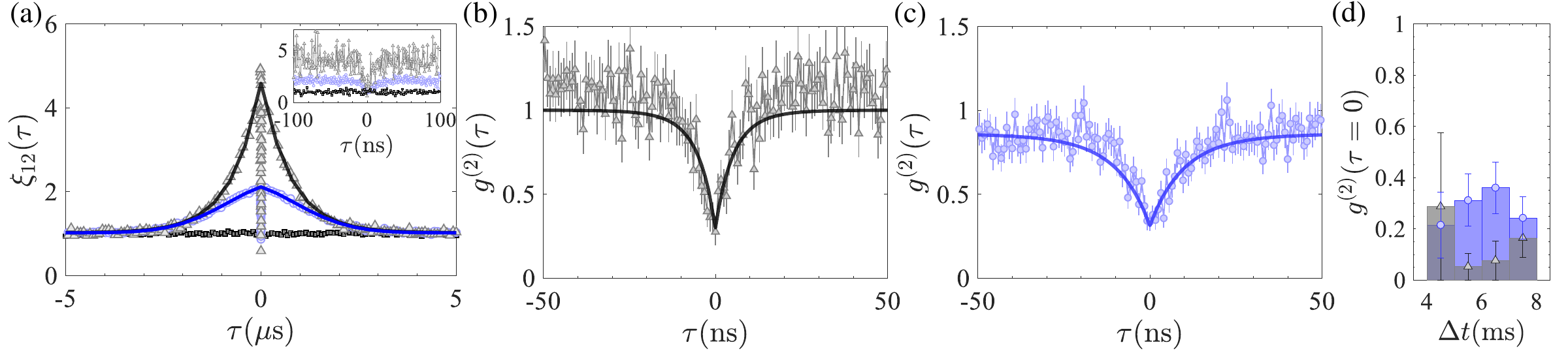}
\caption{\label{fig:fig4} Correlation measurement for the resonant transmission. 
(a) Cross-correlation $\xi_{12}(\tau)$ of two detector counts in the Hanbury-Brown-Twiss setup. Gray triangles (blue circles) show the data obtained with guided atoms in the unplugged (plugged) optical funnel, while black squares show the background $\xi_{12}\approx 1$ obtained without guided atoms. %The central peak in the blue (black) data is showing the average transits for individual atoms with (without) the blue-detuned beam. The gray data shows the reference of the correlation without atomic signal. 
Black and blue lines are bidirectional exponential fits to the data. Inset provides a zoom-in view in the range of $|\tau| \leq 1$00~ns. 
During the detection of atom-transits, normalized intensity correlation functions $g^{(2)}(\tau)$ %shows clear anti-bunching with $g^2(0)=0.27 \pm 0.08$ and $g^2(0)=0.35 \pm 0.08$ 
show clear anti-bunching for (b) unplugged and (c) plugged optical funnels, respectively. Solid lines are theoretical fits \cite{SM}, giving effectively time-averaged single-atom cooperativities $C^+\approx 8$ and 3 for atom-transits in (b) and (c), respectively. Black (blue) bars in (d) show the measured $g^{(2)}(0)$ as a function of guiding time $\Delta t$ in the unplugged (plugged) optical funnel.}
\end{figure*}

To further experimentally characterize guided atoms in the near field, we vary the probe detuning $\Delta \nu$ and measure the transmission spectra at $\Delta t = 5\,\rm{ms}$ %\added{ with and without the repulsive barrier }
[Fig.~\ref{fig:fig3}(c)]. 
With the repulsive barrier, it appears that the transmission spectrum is slightly asymmetric and is red-shifted by $\approx 5\,$MHz from freespace resonance ($\Delta \nu = 0$). This is attributed to the position-dependent light shift induced by the blue-detuned beam; the closer the atom is to the surface, the larger the red-shift. We note that there is negligible light shift from the guiding beam, because $\lambda_\mathrm{r}$ is near the magic wavelength for D2 transition \cite{twocolor2012}. We measure a peak transmission $T\approx 0.26$ and a broad linewidth $\approx 30\,\mathrm{MHz} > \gamma/2\pi$. This can be compared to the case without a repulsive barrier, which gives a symmetric lineshape with smaller peak $T\approx 0.15$ but an even broader linewidth $\approx 37\,\mathrm{MHz}$. We can attribute the reduced transparency in the unplugged funnel to the shorter interaction time $t_i^\mathrm{b} < t_i$ per atom transit. However, the larger linewidth could result from an increased cooperativity and a larger Purcell broadening $\approx (1+\bar{C})\gamma$, where $\bar{C}=\bar{C}^+ + \bar{C}^-$ is the total cooperativity for coupling to both CW-WGM and CCW-WGM. $\bar{C}^- \approx 0.5\bar{C}^+$ for an unpolarized level scheme in Fig.~\ref{fig:fig3}(c) \cite{SM}.

The observed transparency is induced by a continuous stream of atoms interacting with the microring. To extract the guided atom flux in the optical funnel, we fit the measured spectra by calculating a time-averaged transmission signal
\begin{equation}
\frac{T}{T_0} = 1+ \mathcal{N}\left[\frac{\braket{\int \mathcal{T}(\Delta \nu,g(t)) dt}}{\int \mathcal{T}(\Delta \nu,0) dt}-1\right] \,,\label{eq_t}
\end{equation}
where $\mathcal{T}(\Delta \nu,g)$ is the steady-state transmission \cite{SM}, $\braket{...}$ denotes averaging over trajectories calculations as shown in Fig.~\ref{fig:fig3}(a-b), and we have taken into account time- and state-dependent atom-photon coupling as well as light shifts. This model fits well to the measured line shape, using the flux $\mathcal{N}$ as the only adjustable parameter. In either case in Fig.~\ref{fig:fig3}(c), we find $\mathcal{N}\approx 600 \,\rm{ms^{-1}}$, giving near unity $\mathcal{N}t_i^\mathrm{(b)} \approx 0.5- 1.2$. This suggests we have nearly continuous single-atom transits during the entire 1~ms probe window.

We have also measured transmission spectrum using spin-polarized atoms in the $\ket{F=4, m_F=4}$ ground state [Fig. \ref{fig:fig3}(d)], which would have the largest coupling to the CW WGM due to the $\sigma^+$ cycling transition. The spectrum indeed shows a broader linewidth ($\approx 40$~MHz), in accordance with a larger peak cooperativity $C=C^+\approx 8.2$ with peak $g_\mathrm{max}/(2\pi) \approx 136\,$MHz for the $\sigma^+$ transition. Here the fitted atom flux is reduced to $\mathcal{N}\approx 240\,$ms$^{-1}$, likely due to the loss of guided atoms during the optical pumping process. 

To see if the transmitted photons are indeed routed by single atoms one at a time, we perform Hanbury-Brown-Twiss correlation measurements \cite{2008Science_Kimble} on resonant transmissions with polarized atoms. 
In order to do this, the transmitted photon stream is directed from the bus waveguide to an optical fiber with $\gtrsim80\%$ efficiency \cite{3Dtaper} and then detected by two single-photon counters following a 50/50 beam splitter. 
We calculate the intensity cross-correlation by 
\begin{equation}
\xi_{12}(\tau)= \left<\frac{ \overline{I_1(\Delta t)I_2(\Delta t+\tau)}}{\overline{I_1(\Delta t)} \cdot \overline{I_2(\Delta t+\tau)}}\right>\,,\label{eq_xi}
\end{equation}
where $I_{1,2}(t)$ is the time-stamped photon counts from each detector using a $0.8$~ns time bin, $\bar{.}$ and $\braket{...}$ denote averaging over time $\Delta t$ (within a $2$~ms window) and repeated experiments, respectively. The measured $\xi_{12}(\tau)$ shows a peak in the microsecond-timescale with guided atoms in Fig.~\ref{fig:fig4}(a), indicating positive classical correlations in transmitted photons during atom-transits through the evanescence region of the WGM. Using a bidirectional exponential fit, we extract the full width to be $2.4~\mu$s ($1.0~\mu$s) with (without) the repulsive barrier, which is in good agreement with the simulated atom-transit time in Fig.~\ref{fig:fig3}(b). The larger peak correlation measured with the unplugged funnel qualitatively reflects the larger photon scattering rate $\sim C^+\gamma$ during the atom transit. Most importantly, there is a sharp reduction of photon correlations near the central 20~ns window, suggesting the presence of one photon affects the transport of another near the time scale of atom-photon interactions -- similar to a photon-blockade effect \cite{2008Science_Kimble, 2005Nature_Kimble}.
Nonetheless, $\xi_{12}(0)$ does not dip below the shot-noise level, $\xi_{12}(0)= 1$, because the residual classical photon correlation is due to the stochastic nature of randomly arriving single atoms and the finite $\sim 1\%$ transmission of the uncoupled microring resonator. 

To confirm single atom-photon coupling, we extract possible non-classical photon correlations during each detection of atom-transit. We first identify atom-transit events in the time-stamped signals by imposing a threshold of 2 counts within a $1.6~\mu$s running window. We then analyze the normalized intensity correlation $g^{(2)}(\tau)$, similar to Eq.~(\ref{eq_xi}), but using signals in a $2~\mu$s window centered around each post-selected transit events. For details, see the Supplemental Material \cite{SM}. 
In Fig.~\ref{fig:fig4}(b-c), we indeed observe significant photon anti-bunching $g^{(2)}(0)= 0.27(8)$ and $0.35(5)$, respectively, without and with the repulsive barrier. Photon anti-bunching in the resonant transmission signal can be regarded as a signature of single atom coupling to the WGM photons \cite{2008Science_Kimble}. The observed stronger anti-bunching signal in an unplugged funnel again results from a larger time-averaged cooperativity $C^+ \approx 8$, which we determined from a theory fit \cite{SM}.
We also confirm that photon anti-bunching can be observed over an extended time period whenever there are guided atoms coupled to the microring [Fig.~\ref{fig:fig4}(d)]; $g^{2}(0)$ remains nearly at a constant level, including the time around $\Delta t\approx 5~$ms when we observe the peak atom flux. This suggests the transmitted photons observed in Figs.~\ref{fig:fig2}-\ref{fig:fig3} are indeed routed by single atoms, one at a time, instead of multiple atoms simultaneously coupled to the same resonator mode, in which we expect more complex behavior in photon correlations. Nonetheless, correlated photon transport gated by multiple atoms is an interesting topic in its own right \cite{2016Science_Alp, 2023PRX_Stanford}. This can be studied using multiple optical funnels formed on a microring resonator.

In conclusion, we have demonstrated an optical trapping technique that guides single atoms on a planar nanophotonic resonator with subwavelength precision. Using our technique, single atom trapping probability may be improved by pulsing on a lattice beam \cite{2013Nature_Lukin, PRL2015_Superradiance, 2019NC_MayKim, 2019PRL_nanoFiberCavity_Nayak, 2020PRL_Rauschenbeutel_WGM} to localize atoms in the near field following an instantaneous feedback from probing a WGM resonator \cite{2020PRL_Rauschenbeutel_WGM}. 
To further cool and localize single atoms in a near field trap, evanescent-wave cooling \cite{1996PRA_evanescentCooling, 1997PRL_evanescentCooling}, Raman sideband cooling \cite{2018PRX_Rauschenbeutel_cooling, 1998PRL_dRSC_Chu}, or cavity cooling \cite{2004Nature_cavitycooling,2017PRL_cavitycooling} may be implemented. The achieved single-atom, single-mode (CW-WGM) cooperativity $C^+\gtrsim 8$ is currently limited by the quality factor $Q\approx 2\times 10^5$ of the coupled microring circuit. 
We expect significant improvement in the cooperativity parameter by more than 5-fold with a better $Q>10^6$ \cite{2019Optica_Chang,2021SPIE_Hikaru} following improvements in waveguide surface roughness and material quality. Our work would enable new applications, for example, in chiral quantum optics \cite{2017Nature_Chiral,2014Science_Dayan,2016Science_Circulator,2018NaturePhysic_Dayan, 2017PRL_ChiralQuantumNetworks, 2015PRB_ChiralEntanglement, 2015PRA_chiralSpinNetworks} based on cold atoms coupled to an on-chip WGM resonator. 
Our system also holds a promise for realizing photon-mediated atom-atom interactions and quantum many-body physics \cite{2015NaturePhotonics_manybody,2015NaturePhotonics_subwavelength,2011PRL_Lev_proposal,2016PNAS_SpinModel,2019PRL_Monika_photon_mediated_interaction, 2019PRL_Lev_photon_mediated_interaction} with multiple trapped atoms. 

We thank B. M. Fields and M. E. Kim for their prior contribution to the experiment. This work was supported by the AFOSR (Grant NO. FA9550-17-1-0298, FA9550-22-1-0031), the ONR (Grant NO. N00014-17-1-2289), and a seed fund from Purdue Quantum Science and Engineering Institute. H.T. and C.-L.H. acknowledge support from the NSF (Grant NO. PHY-1848316).

\bibliography{apssamp}

\begin{thebibliography}{75}
\bibitem [{not()}]{note_sim}%
  \BibitemOpen
  \href@noop {} {\bibinfo {title} {{We perform trajectory simulations in the
  $x$-$z$ plane. Motion along the $y$-axis is less relevant due to symmetry of
  the waveguide, the large dipole beam waist $7\,\mu$m, and its long Rayleigh
  length $z_\mathrm{R}\approx 165\,\mu$m.}}}\BibitemShut {Stop}%
\end{thebibliography}

\end{document}

\setcounter{page}{8}
\thispagestyle{plain}

\title{Supplemental Material for:\\Coupling single atoms to a nanophotonic whispering-gallery-mode resonator via optical guiding}
\appendix
%\date{\today }
\maketitle

\section{Experimental setup}
The Si$_3$N$_4$ microring resonator used in this experiment is fabricated on a transparent SiO$_2$-Si$_3$N$_4$ double-layer membrane, suspended over a large window (2 mm $\times$ 8 mm) on a silicon chip. The open window ensures optical access for conventional laser cooling above the chip and freespace optical guiding adopted in this study. The thickness of the SiO$_2$ layer is $\approx 2.04\,\mathrm{\mu m}$ while the bottom Si$_3$N$_4$ layer is $\approx 583\,\mathrm{nm}$ thick. As shown in Fig.~1, the microring waveguide width and height are $\approx 950\,\mathrm{nm}$ and $\approx 326\,\mathrm{nm}$, respectively. The waveguide forms a racetrack-shaped ring resonator, with $15\,\mu$m bend radius and one 8$\,\mu$m-long straight part in each side. Microring waveguide circumference is $\approx 110\,\mu$m. A bus waveguide couples to the microring in a pulley-shaped coupling region and is end-coupled to two cleaved fibers for photon input and output (with at least one port approaching $\gtrsim 80\,\%$ single-pass coupling efficiency \cite{3Dtaper}). The cleaved fibers are pre-aligned and are epoxied in U-shaped grooves fabricated on the silicon chip. Details about the design and fabrication processes of the microring circuit and the fiber couplers can be found in Ref. \cite{2020APL_Chang,3Dtaper}.

\begin{figure}[!h]
\centering
\includegraphics[width=1\columnwidth]{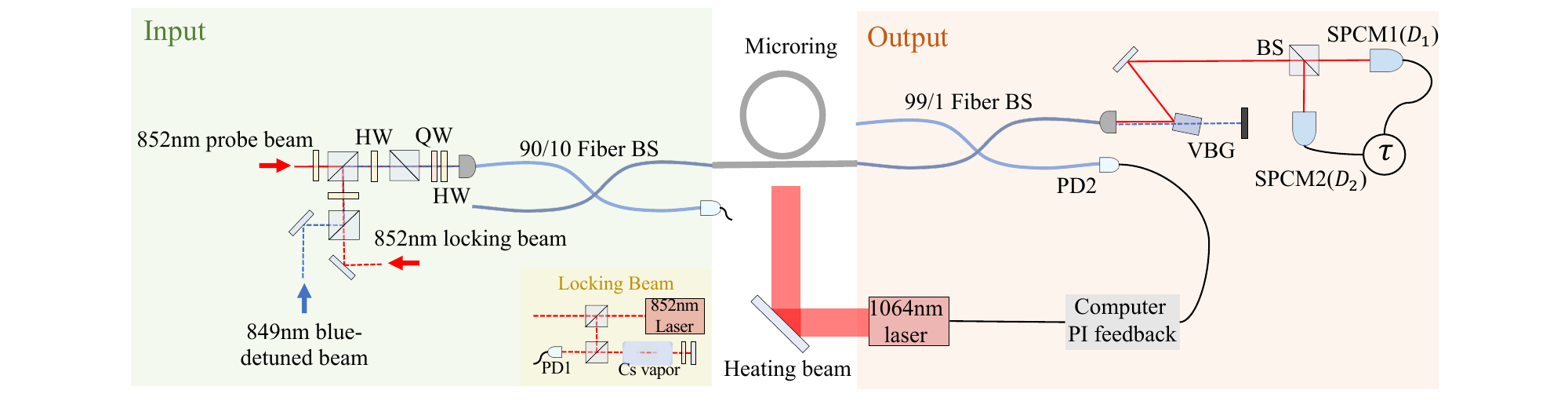}
\caption{\label{fig:figS1} Experimental setup for probing atom-microring coupling.  }
\end{figure}

The experimental setup for on-chip laser cooling is essentially identical to that described in Ref.~\cite{2019NC_MayKim}, except the majority of polarization gradient cooled (PGC) atoms are located $\sim 250~\mu$m above the chip surface due to slightly different alignment of the cooling beams. The experimental setup for probing interaction between cold atoms and a microring resonator is shown in Fig.~\ref{fig:figS1}. After the microring circuit is integrated into the ultrahigh vacuum science chamber, 
optical fibers are guided out of the chamber via teflon fiber feedthroughs \cite{1998_feedthrough} and are connected to the input and output fiber stages.
The input fiber stage accepts three beams (852~nm probe beam, 852~nm locking beam and 849~nm blue WGM beam), whose polarizations are carefully adjusted via half- and quarter-waveplates (HW and QW) to excite transverse-magnetic-like whispering-gallery modes (TM WGMs). The 849~nm beam runs continuously while the 852~nm probe and locking beams are pulsed on at different times. A 90:10 fiber beam splitter (BS) is inserted in the input stage for intensity pickup and monitoring. 
At the output, a volume Bragg grating (VBG) is utilized to filter out the 894~nm component, so that the 852~nm probe photons can be recorded either using a single photon counting module (SPCM) or by two SPCMs in a Hanbury-Brown-Twiss (HBT) setup as in Fig.~\ref{fig:figS1}. 
A 99:1 fiber BS picks up $1\%$ of the transmitted light for monitoring the 852~nm locking beam, whose frequency is referenced to a cesium vapor cell.

We illuminate the circuit with a 1064~nm external heating beam (up to 300~mW) to thermally tune the microring resonances. Frequency of the probe TM-WGM is thermally stabilized to the $F=4\leftrightarrow F'=5$ transition via a computer-controlled feedback that monitors the transmission of the locking beam. 
In each experiment, we pulse on the locking beam immediately following the probe pulses so to stabilize the WGM resonance without affecting the atom-microring coupling. 
Figure~\ref{fig:figS2}(a) shows the transmission spectrum across the probe WGM resonance, where the line center is well-stabilized to within $\pm12\,$MHz relative to the atomic resonance. The frequency error is around $0.6\,\%$ of the WGM resonance linewidth, and has negligibly small effect on probing atom-WGM interaction.

Wavelength of the blue WGM, $\lambda_\mathrm{b}=849.55~$nm, is one free spectral range away from the probe WGM. 
A small blue-detuned beam power ($33\,\mu$W) is used to excited the blue WGM (at an equivalent circulated power of $3.2\,$mW) to create a strong repulsive potential near the microring surface; see later discussions. We excite the blue WGM continuously to avoid any transient thermal effect, and keep the probe WGM frequency stabilized during the experiment. 

\section{Characterization of the microring resonator}
\begin{figure}[!h]
\centering
\includegraphics[width=1\columnwidth]{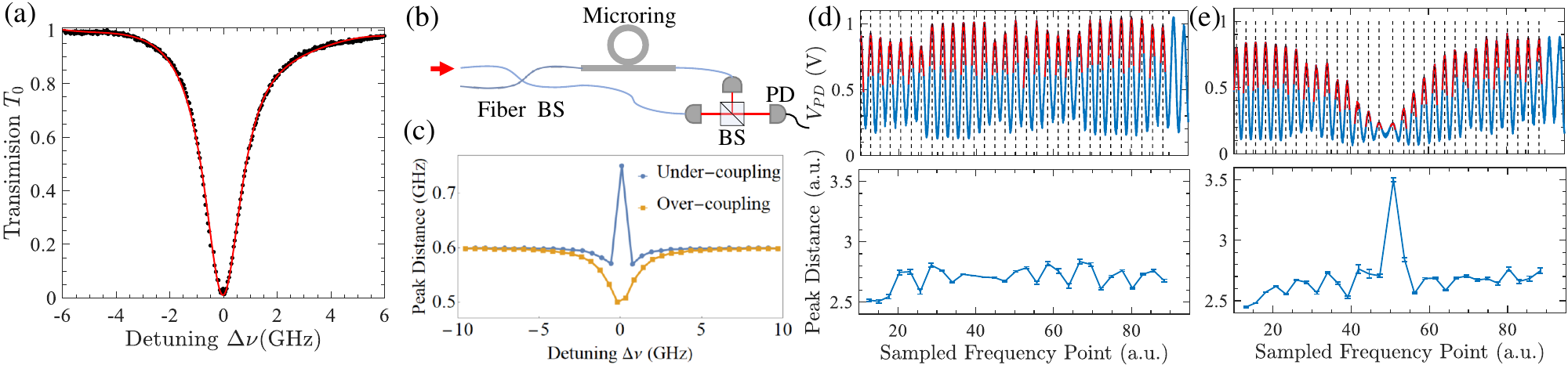}
\caption{\label{fig:figS2} Characterization of the microring resonator. (a) Transmission spectrum of the bare microring resonator. (b) Interferometer setup for determining the bus waveguide coupling condition. (c) Theoretical prediction of the frequency period of the interferometer for different coupling conditions. (d-e) Experiment data when the sweeping range of the laser frequency does not overlap (d) or overlaps (e) with the microring resonance, respectively. 
}
\end{figure}

We design the bus waveguide pulley-coupler to achieve near critical-coupling condition, where the waveguide coupling rate $\kappa_{\mathrm{e}}$ equals the intrinsic photon loss rate $\kappa_{\mathrm{i}}$. In reality, a small mismatch would exist and a fiber Mach-Zehnder (MZ) interferometer [Fig.~\ref{fig:figS2}(b)] is utilized to determine whether the microring is under-coupled ($\kappa_{\mathrm{e}} < \kappa_{\mathrm{i}}$) or over-coupled ($\kappa_{\mathrm{e}} > \kappa_{\mathrm{i}}$). In the latter case, a $\pi$ phase shift in the bus waveguide transmission would result, when the probe frequency is swept across a microring resonance. In an unbalanced MZ interferometer with slightly longer arm in the microring path, fringe periodicity will increase (decrease) for under-coupled (over-coupled) condition [Fig.~\ref{fig:figS2}(c)]. Figure~\ref{fig:figS2} (d) and (e) show the photodetector reading V$_{\mathrm{PD}}$ of the interferometer output and the extracted peak distances as a function of probe frequency. Comparing with the output far from the microring resonance (d), 
a clear increase of peak distance is observed in (e) near the microring resonance. This indicates that the microring is in the under-coupled condition. 

We fit the transmission spectrum of the microring resonator [Fig.~\ref{fig:figS2}(a)], knowing that $\kappa_{\mathrm{e}}<\kappa_{\mathrm{i}}$. We extract the external coupling rate $\kappa_{\mathrm{e}}\approx 2\pi \times0.77\,$GHz and the intrinsic loss rate $\kappa_{\mathrm{i}}\approx2\pi \times0.95\,$GHz. The total photon loss rate is $\kappa_{\mathrm{tot}}\approx2\pi \times 1.72\,$GHz, giving a quality factor $Q\approx 2\times 10^5$ for the probe WGM resonance studied in the experiment. 

\begin{figure}[!h]
\centering
\includegraphics[width=0.5\columnwidth]{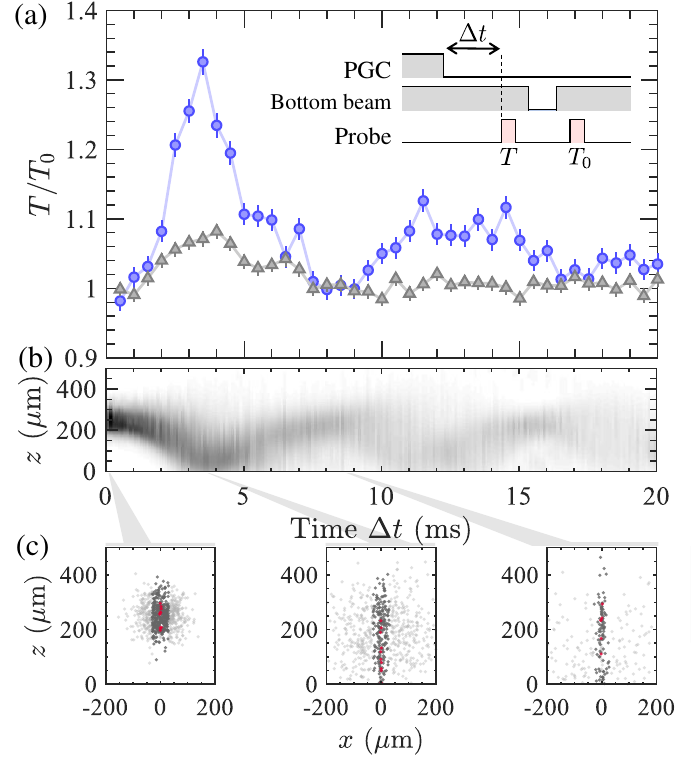}
\caption{\label{fig:figS_guiding} Sample atom guiding signal in the optical funnel. (a) Sample resonant transmission $T/T_{0}$ signal versus guiding time $\Delta t$ in a plugged (circles) and unplugged (triangles) funnel, respectively. %Experimental sequence is illustrated in the inset. 
(b) Normalized atomic distribution along the $z$ axis, evaluated by Monte Carlo trajectory calculations using the experimental initial conditions. 
(c) Sample trajectory distributions at the indicated timing. Dark gray (light gray) circles mark those initially loaded (not loaded) in the funnel potential, while red circles label those entering the near field. }
\end{figure}

\section{Atom guiding in the optical funnel}
In Fig.~\ref{fig:figS_guiding}(a), we present sample resonant transmission data obtained from a resonator other than the one used in the main text. Similarly, there is an increased transmission and a clear maximum at $\Delta t \approx 3.5\,\mathrm{ms}$, indicating a peak atom flux arriving at the near field. A second transmission peak can be found at $10\,{\mathrm{ms}} \lesssim \Delta t\lesssim 16\,$ms. Similar oscillatory behavior can also be found in atomic fluorescence with a period of $\sim 9\,\mathrm{ms}$ (see discussions in the next section), roughly matching half of the longitudinal trap period of the bottom beam potential. This suggests guided atoms are reflected longitudinally and are drawn back towards the surface for recoupling. 
For atom guiding without the repulsive barrier, observed transmission increase is less pronounced. However, oscillations in atomic fluorescence remain visible.

We find that atoms reflect longitudinally in the funnel for two main reasons. One is apparently due to the repulsive barrier in the plugged funnel. The other one arises from rapid tapering of the guiding potential within $z\lesssim 20\,\mu$m [Fig.~\ref{fig:figS3}(a)]. When a guided atom experiences strong transverse compression, it would quickly gain transverse kinetic energy while reducing the longitudinal kinetic energy according to the classical adiabatic theorem \cite{2002PRA_adiabatic_theorem,2011PRA_funnel}. This results in longitudinal deceleration that can even reflect an atom before it approaches the near field. 

Our atomic trajectory simulations \cite{note_sim} confirm the observed guiding effect. In Fig.~\ref{fig:figS_guiding}(b), simulated density near the waveguide surface ($z\approx 0$) shows quasi-periodic oscillations that align well to the observed oscillatory peaks in Fig.~\ref{fig:figS_guiding}(a). 
Within those guided trajectories, $\sim 6\,\%$ near the center ($x\approx 0$) and with a small initial transverse velocity can be directed to the near field [Fig.~\ref{fig:figS_guiding}(c)]. 
Compared with free-falling cases, the number of atoms arriving on the waveguide is estimated to increase by $40$ times with optical guiding.

\section{Fluorescence imaging longitudinal bounces in the optical funnel}
\begin{figure}[!h]
\centering
\includegraphics[width=1\columnwidth]{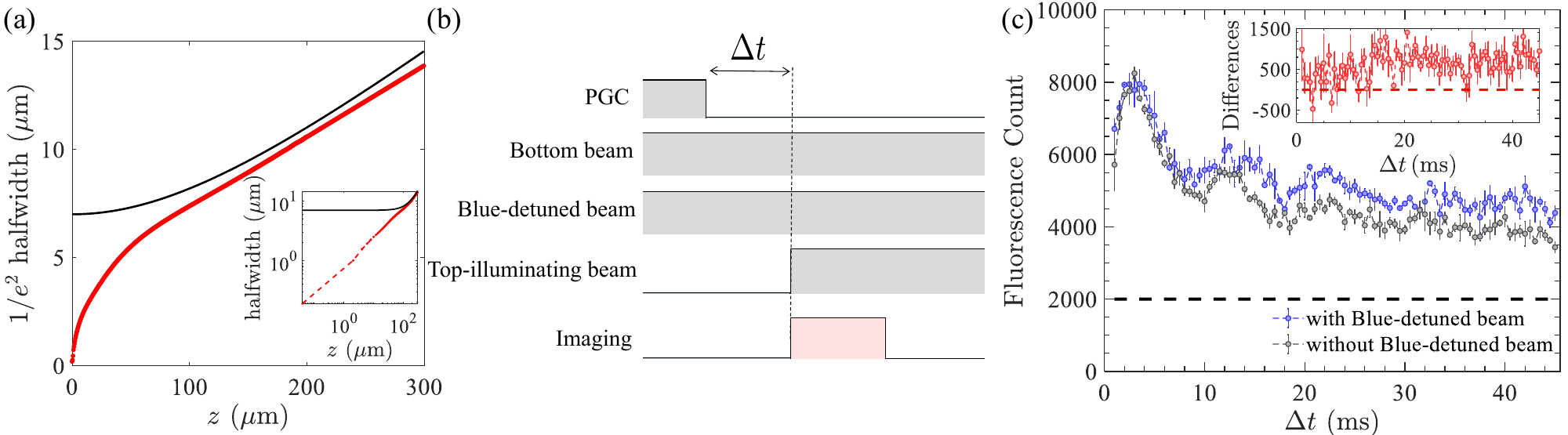}
\caption{\label{fig:figS3} Analysis of the optical funnel and dynamics of guided atoms observed from fluorescence imaging. (a) $1/e^2$ halfwidth of the optical funnel (red), comparing with the width of the same beam propagating in free space (black curve). Inset shows the plot in logarithmic scale. (b) Experiment procedure for fluorescence detection. 
(c) Fluorescence counts as a function of waiting time $\Delta t$. Dashed line marks the background. Inset shows the difference in atomic fluorescence with and without the repulsive barrier. }
\end{figure}

Fluorescence imaging provides complementary information on the dynamics of guided atoms in the optical funnel. 
The imaging procedure is shown in Fig.~\ref{fig:figS3}(b). 
After polarization gradient cooling (PGC) and a wait time $\Delta t$, we switch on a top-illuminating beam with $7\,\mu$m beam waist to freeze out longitudinal atomic motion in a tight lattice potential and perform fluorescence imaging \cite{2019NC_MayKim}. Atomic fluorescence is collected by a microscope objective (numerical aperture NA$=0.35$), and is wavelength- as well as polarization-filtered to remove scattered light from the microring waveguide. We record the image on an electron multiplication charge-coupled device, % for $30\,\rm{ms}$. 
and monitor the fluorescence counts within a 4-by-4 pixels box ($3\times3\,\mu$m$^2$ area), as shown in Fig.~1(c). Depth of focus of the objective is $\sim \pm 10~\mu$m and the imaged atomic fluorescence mainly comes from atoms captured at $z\lesssim 10\,\mu$m above the microring waveguide.

Measured fluorescence counts versus wait time $\Delta t$ is shown in Fig.~\ref{fig:figS3}(c). 
Similar to the observed transmission peak in Fig.~\ref{fig:figS_guiding}(a), atomic fluorescence clearly shows a maximum at $\Delta t\approx3.5\,$ms. Within this hold time, the optical funnel draws the atomic cloud towards the surface at $z\lesssim 10\,\mu$m, increasing the number of imaged atoms within the depth of focus. We also note that there is already atomic fluorescence detected at $\Delta t \approx 0\,$ms. Whereas, there is no clear initial signal in probe transmission as shown in Fig.~2. This suggests we have low atom loading efficiency in the near field region during PGC even with the presence of trapped atoms around $z\lesssim 10\,\mu$m in the optical funnel potential.

In addition, atomic fluorescence shows multiple peaks regardless of the presence of a repulsive barrier. 
This suggests that atoms can bounce off longitudinally within the optical funnel multiple times, and can be guided back towards the surface periodically. 
Indeed, longitudinal reflection is anticipated in an optical funnel due to transverse atomic motion in a strongly tapered guiding potential \cite{2002PRA_adiabatic_theorem}, as discussed in Appendix~C. Figure~\ref{fig:figS3}(a) shows rapid narrowing of the optical funnel within $z\lesssim 20\,\mu$m, where the transverse width of the funnel rapidly drops below $1\,\mu$m and reaches $200~$nm $1/e^2$-halfwidth on the waveguide surface. Our Monte Carlo trajectory calculations confirm that most guided atoms bounce off in this region prior to arriving at the near field. Only those with small transverse kinetic energy and displacement can enter the near field region. Approaching the waveguide surface, these atoms can still be reflected by the repulsive barrier, if present. Otherwise, they will be attracted to the surface due to Casimir-Polder interactions. Comparing measured fluorescence with and without the repulsive barrier, we find that barrier-reflected atoms are responsible for the increased counts after the first peak (or the first bounce) in Fig.~\ref{fig:figS3}(c). 

\section{Light Shift Calculation and the Casimir-Polder potential}
The optical potential experienced by atoms can be separated into contributions from scalar (rank-0), vector (rank-1) and tensor (rank-2) terms in the irreducible tensor representation
\begin{equation}
  U(\mathbf{r})=U^s(\mathbf{r})+U^v(\mathbf{r})+U^t(\mathbf{r})
\end{equation}
where
\begin{align}
      U^s(\mathbf{r}) = & -\alpha^{(0)}(\omega)|\mathbf{E}(\mathbf{r})|^2\\
      U^v(\mathbf{r}) = & -i\alpha^{(1)}(\omega)\frac{\mathbf{E}(\mathbf{r})\times \mathbf{E}^*(\mathbf{r})\cdot\hat{\mathbf{F}}}{2F}\\
      U^t(\mathbf{r}) = & -\alpha^{(2)}(\omega)\frac{3}{F(2F-1)} \left[\frac{\hat{F}_\mu \hat{F}_\nu + \hat{F}_\nu \hat{F}_\mu}{2} -\frac{\hat{\mathbf{F}}^2}{3}\delta_{\mu\nu}\right]E_\mu E^*_\nu
\end{align}
and $\alpha^{(0,1,2)}(\omega)$ are the corresponding scalar, vector, and tensor polarizabilities, $\hat{\mathbf{F}}$ is the total angular momentum operator, $F$ is the quantum number, and $E_{\mu}$ ($\mu\in\{x,\,y,\,z\}$) are the electric field components in Cartesian coordinates. We calculate the dynamical polarizabilities using transition data summarized in \cite{2013_Lightshift_ref}. TABLE \ref{alphatable} lists the value of polarizabilities used in the calculation of the trap potential in the main text.

\begin{table}[!h]
    \centering
    \begin{tabular}{c|c|c|c|c|c|c}
        $\lambda$ & \multicolumn{3}{c|}{$\lambda_\mathrm{r}$} & \multicolumn{3}{c}{$\lambda_\mathrm{b}$}\\  \hline
        {} & $\alpha^{(0)}$ (a.u.) & $\alpha^{(1)}$ (a.u.) & $\alpha^{(2)}$ (a.u.)  & $\alpha^{(0)}$ (a.u.) & $\alpha^{(1)}$ (a.u.) & $\alpha^{(2)}$ (a.u.)\\  \hline
         6S$_{1/2}\,F=4 $ & 3025 & -1625  & 0 & -39046 & -35409  & 0\\
         6P$_{3/2}\,F'=5 $ & 3025 & 3556  & 579 & 18077 & 55643  & -18785\\
    \end{tabular}
    \caption{Cesium 6S$_{1/2}, F=4$ ground state and 6P$_{3/2}, F'=5 $ excited state dynamic polarizabilities at $\lambda_\mathrm{r} =935.3~$nm and $\lambda_\mathrm{b}=849.55~$nm.}
    \label{alphatable}
\end{table}

\begin{figure}[!h]
\centering
\includegraphics[width=1.0\columnwidth]{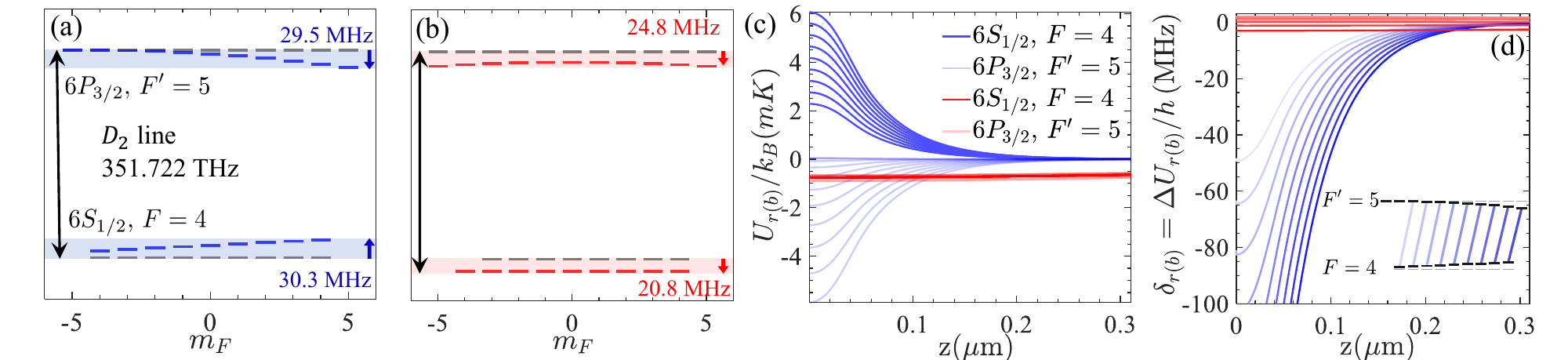}
\caption{\label{fig:figS4} Total light shifts (sum of scalar, vector and tensor light shifts) caused by (a) the blue WGM ($\lambda_\mathrm{b}=849.55~$nm, $\sigma^+$ polarization) and (b) the bottom beam ($\lambda_\mathrm{r} =935.3~$nm, linearly polarized), with $1\,$mK (20.8~MHz) shift for the $m_F=0$ ground state. Blue (red) lines show the light shifts of different $m_{F}$ states and gray lines mark the unperturbed states respectively. Shade bands mark the maximum total light shift. (c) Potential linecuts $U_{r}(x=0,z)$ (red curves) and $U_{b}(x=0,z)$ (blue curves) calculated using powers $P_{r}=15\mathrm{mW}$ and $P_{b}=33\mathrm{\mu W}$ in the bottom beam and in the blue-detuned beam, respectively. (d) Differential light shifts $\delta_{r,b}(0,z)$ between the indicated states in the inset, caused by the same bottom beam (red curves) and the `blue' WGM (blue curves) as in (c), respectively.}
\end{figure}

Relevant light shifts for different Zeeman sub-levels in the $6S_{1/2}$ ground state and the $6P_{3/2}$ excited state are shown in Fig.~\ref{fig:figS4}. 
The red-detuned, bottom-illuminating beam (wavelength $\lambda_\mathrm{r}=935.3~$nm) has linear polarization parallel to the illuminated microring waveguide ($8\,\mu$m straight part of the racetrack), and primarily induces scalar light shift in the ground state, and scalar plus weak tensor light shifts in the excited state; see Fig.~\ref{fig:figS4}(b). Moreover, $\lambda_\mathrm{r}=935.3~$nm is a magic wavelength for cesium D2 transition, the differential light shifts between ground and excited states are only caused by small tensor shifts; see Fig.~\ref{fig:figS4}(c-d).
The blue WGM (wavelength $\lambda_\mathrm{b}=849.55~$nm) has a locally defined $\sigma^{+}$ polarization [see Fig.~\ref{fig:figLevel}(a)]. It additionally induces vector light shifts in both the ground and the excited states; see Fig.~\ref{fig:figS4}(a). For guided atoms in a plugged optical funnel, differential light shifts are mainly caused by the blue WGM [Fig.~\ref{fig:figS4}(d)]. 

In Fig.~1(d), we plot the total potential of a plugged funnel while considering the effect of atom-surface Casimir-Polder interactions. Here, we approximate the ground state Casimir-Polder (CP) potential with $U_{\mathrm{CP}} = -{C_{4}}/{z^{3}(z + \lambdabar)}$, where $z$ is atom-surface distance, $C_{4}/h = 267\,\mathrm{Hz}\cdot\mathrm{\mu m}^4$, $h$ is the Planck constant and $\lambdabar = 136\, \mathrm{nm}$~\cite{2019NC_MayKim}. In our trajectory calculations, most guided atoms experience a classical turning point at $z\gtrsim 100~$nm in a plugged potential. The resulting CP shift is $\lesssim h\times 1~$MHz and is small compared with light shifts created by the optical funnel. 

\section{Theoretical Models}
\setcounter{equation}{4}
\subsection{Bus waveguide-microring coupling}
To model the microring system, we start from the fundamental modes in microring resonator using coupled-mode theory, see details in supplementary material of Ref. \cite{2019Optica_Chang}. We denote the amplitude of two fundamental transverse-magnetic (TM) modes as $a$ for clockwise (CW) mode and $b$ for counter-clockwise (CCW) mode. The resonator field can be decomposed into these two counter-propagating modes of interest $\mathbf{E}(r,t)=a\mathbf{E}_{+}(r,t)+b\mathbf{E}_{-}(r,t)$, where $\mathbf{E}_{\pm}(r,t)=\mathbf{E}_{\pm}(r)e^{-i\omega t}$ and
\begin{equation}
    \mathbf{E}_\pm(\mathbf{r}) = \left[ \mathcal{E}_\rho (\rho,z)\hat{\rho} \pm i \mathcal{E}_\phi(\rho,z)\hat{\phi} + \mathcal{E}_z (\rho,z)\hat{z}\right]e^{\pm i m\phi}.\label{appendix:mode:A1}
\end{equation}
Here, $\mathcal{E}_{\mu}(\rho,z) (\mu=\rho,\phi,z)$ are real functions independent of $\phi$ due to cylindrical symmetry. In our experiment, we excite the resonator mode by sending light from one end of the bus waveguide. We have the following coupled rate equations
\begin{align}
    \frac{da}{dt} = & -\left(\frac{\kappa}{2} + i\Delta_c \right)a(t) + i\beta e^{i \xi} b(t) + i\sqrt{\kappa_e}s_{\mathrm{in}}\nonumber\\
    \frac{db}{dt} = & -\left(\frac{\kappa}{2} + i\Delta_c \right)b(t) + i\beta e^{-i \xi}a(t)
\end{align}
where $\Delta_c=\omega_{\mathrm{c}} - \omega$ is the resonator detuning relative to the driving field, $\beta$ is the coherent back-scattering rate and $\xi$ is the phase. 
$s_{\mathrm{in}}$ is normalized amplitude of the input power in the bus waveguide so that $P_{in}=\hbar\omega |s_{\mathrm{in}}|^{2}$. The steady-state mode amplitudes are
\begin{align}
    a = & i\sqrt{\kappa_{e}} \frac{(\frac{\kappa}{2} + i \Delta_c)s_{\mathrm{in}}}{(\frac{\kappa}{2} + i \Delta_c)^{2}+\beta^{2}}\nonumber\\
    b = & i\sqrt{\kappa_{e}} \frac{i \beta e^{-i \xi} s_{\mathrm{in}}}{(\frac{\kappa}{2} + i \Delta_c)^{2}+\beta^{2}}
\end{align}
From our characterization of microring [Fig.~\ref{fig:figS2}(a)], the back-scattering rate $\beta$ is much smaller than the linewidth of the cavity, so we neglect it and set $\beta=0$. In this case, $b = 0$ and we only excite the CW mode with amplitude $a$. The transmission spectrum of a bare microring is then 
\begin{equation}
T_0(\Delta_c)=\left|1+i\sqrt{\kappa_e}\frac{a}{s_\mathrm{in}}\right|^2 = \left| 1- \frac{ 2\kappa_e}{\tilde{\kappa}}\right|^2\,, \label{Eq:SM:bareT}
\end{equation}
where $\tilde{\kappa}= \kappa + 2i \Delta_c$. This is the formula we used for fitting the transmission spectrum of a bare resonator. 

\subsection{Atom-WGM photon coupling}

We now consider atom-WGM photon coupling. In cavity QED, single atom-photon coupling strength is defined as
\begin{equation}
    g_i=d_i\sqrt{\frac{\omega}{2\hbar\epsilon_{0}V_{\mathrm{m}}}}\label{Eq: g_definition}
\end{equation}
where $d_i = \braket{e|\mathbf{d}|g} \cdot \mathbf{e}_i$ is the transition dipole moment, $i$ labels $\sigma^\pm$ or $\pi$ transitions, $\mathbf{e}_i$ is the polarization unit vector, 
$\hbar$ is the reduced Planck's constant, $\epsilon_{\mathrm{0}}$ is the vacuum permittivity, and $V_m(\mathbf{r}_{a})$ is the effective mode volume at atomic position $\mathbf{r}_{a}=(\rho_a, \phi_a, z_a)$,
\begin{equation}
    V_{m}(\rho_{a},z_{a})=\frac{\int{\epsilon(\rho,z)|\mathbf{E}_\pm(\rho,z)|^{2}\rho d\rho dz d\phi}}{\epsilon(\rho_{a},z_{a})|\mathbf{u}_i\cdot\mathbf{E}_\pm(\rho_{a},z_{a})|^{2}}.
\end{equation}

\begin{figure}[t!]
\centering
\includegraphics[width=1.0\columnwidth]{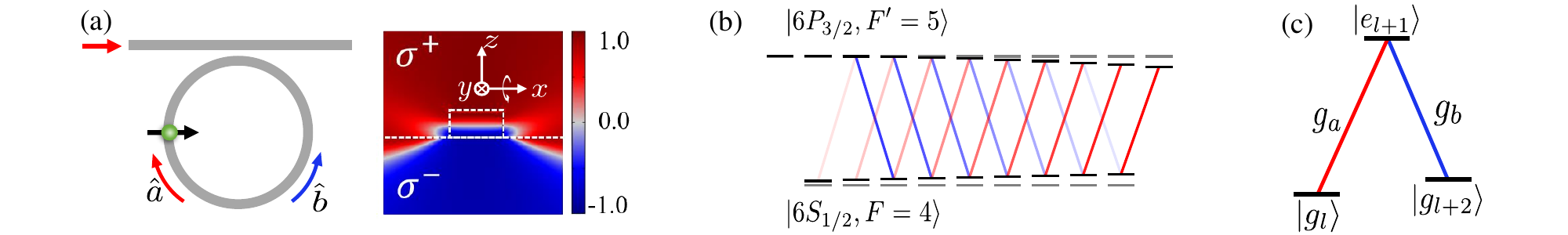}
\caption{(a) Degree of circular polarization of the CW WGM (mode $\hat{a}$) locally around the microring waveguide, calculated as $(|\mathbf{E}_+\cdot\mathbf{e}^*_+|^2-|\mathbf{E}_+\cdot\mathbf{e}^*_-|^2)/|\mathbf{E}_+|^2$, and $\mathbf{e}_\pm = (\mathbf{e}_y \pm i\mathbf{e}_z)/\sqrt{2}$. Note the mode propagation is along the $y$-axis, but the circular polarization is about the $x$-axis. (b) Multilevels of the $\ket{F=4}\leftrightarrow\ket{F'=5}$ transition. (c)~Effective $\Lambda$-level and related coupling ($\ket{g_l}\leftrightarrow \ket{e_{l+1}}\leftrightarrow \ket{g_{l+2}}$). }\label{fig:figLevel}
\end{figure}

As shown in Fig.~\ref{fig:figLevel}(a), the TM WGM is nearly circularly polarized above the waveguide because of strong confinement of the electric field. Polarization rotation of the WGM is also locked to the circulation direction. We can thus consider a case that the atomic spin axis is aligned transversely to the microring waveguide, and CW (CCW) WGM photons would locally drive only the $\sigma^{+}$ ($\sigma^{-}$) transition, as shown in Fig.~\ref{fig:figLevel}. In the case that an atom is spin-polarized to the $\ket{F=4, m_F=4}$ ground state, the atom-CW WGM photon coupling strength $g_a$ can be maximized for the $\ket{F=4, m_F=4} \leftrightarrow \ket{F'=5, m_{F'}=5}$ cycling transition,
\begin{equation}
g=\sqrt{\frac{3\lambda^{3}\omega\gamma}{16\pi^2 V_{\mathrm{m}}}}\,,
\end{equation}
where we have used $d^{2}=\frac{3\lambda^{3}\epsilon_{0}\hbar\gamma}{8\pi^{2}}$, $\lambda$ is the transition wavelength, and $\gamma$ is the atomic decay rate in freespace. Generally, the atom-CW(CCW) WGM photon coupling strength $g_{a} (g_{b})$ for a $\ket{F=4, m_F=l}$ ground state can be written as
\begin{align}
g_{a} &= \eta^+_l g \\
g_{b} &= \eta^-_l g
\end{align}
where $\eta^{\pm}_l = \sqrt{2}C_{l,l\pm1}$ and $C_{l,l\pm1}$ is the dipole matrix element for the $\sigma^{+(-)}$ transition. For example, $\eta^-_4 = 1/\sqrt{45}$ for the $\ket{F=4, m_F=4} \leftrightarrow \ket{F'=5, m_{F'}=3}$ transition. 

In Fig.~3(a), we plot the averaged $\bar{g}$ of all magnetic sub-levels for the atom-WGM photon coupling. Here $\bar{g}^2 \equiv \bar{g}_a^2 =\bar{g}_b^2$, and the averaged cooperativity $\bar{C} \equiv 4\bar{g}^2/\kappa/\gamma = \bar{C}^+ = \bar{C}^-$, where $C^{+(-)} = 4g_{a(b)}^2/\kappa/\gamma$ is the cooperativity for atom-CW (CCW) WGM photon coupling, and $\bar{.}$ denotes averaging over all magnetic sub-levels.

\subsection{Transmission spectrum of coupled atom-microring-bus waveguide system}
We calculate the transmission spectrum in the weak driving limit, by solving for the transmission coefficient of single photon transport in the atom-coupled microring system. For completeness, we consider a multi-level atom coupled to both of the CW and CCW WGMs, and adopt a real-space representation for discussing single photon transport in the system \cite{2009PRA1_ShanHuiFan,2009PRA2_ShanHuiFan,2018PRA_FrancoNori}. Including dissipation in the WGMs and the atomic excited states, we consider a non-Hermitian Hamiltonian in the quantum-jump approach~\cite{StatisticalMethodsInQuantumOptics2}:
\begin{align}
    \hat{H}_{\mathrm{eff}}/\hbar= & \left(\omega_{c}-i\frac{\kappa_{i}}{2}\right)\left(\hat{a}^{\dagger}\hat{a}+\hat{b}^{\dagger}\hat{b}\right)+ \int dx\,\hat{c}_{\mathrm{R}}^{\dagger}(x)\left(\omega_{0}-iv_{\mathrm{g}}\frac{\partial}{\partial x}\right)\hat{c}_{\mathrm{R}}(x)+
    \int dx\hat{c}_{\mathrm{L}}^{\dagger}(x)\left(\omega_{0}+iv_{\mathrm{g}}\frac{\partial}{\partial x}\right)\hat{c}_{\mathrm{L}}(x)\nonumber\\
    & +\int dx \delta(x)\left[V_{a}\hat{c}_{\mathrm{R}}^{\dagger}(x)\hat{a}+V_{a}^{*}\hat{c}_{\mathrm{R}}(x) \hat{a}^{\dagger}\right]
    +\int dx \delta(x)\left[V_{b}\hat{c}_{\mathrm{L}}^{\dagger}(x)\hat{b}+V_{b}^{*}\hat{c}_{\mathrm{L}}(x)\hat{b}^{\dagger}\right]\nonumber\\
    &  +\sum_l \left(\Omega_{e_l}-i\frac{\gamma}{2}\right)\ket{e_l}\bra{e_l}+ \sum_l \Omega_{g_l}\ket{g_l}\bra{g_l} 
    +g\sum_l \left[\eta^+_{l}\left(\hat{a}\hat{\sigma}_{+,l}^{\dagger}+\hat{a}^{\dagger}\hat{\sigma}_{+,l}\right)+\eta^-_l\left(\hat{b}\hat{\sigma}_{-,l}^{\dagger}+\hat{b}^{\dagger}\hat{\sigma}_{-,l}\right)\right]\,, \label{EqSM:sys}
\end{align}
where the terms in the first two lines describe the bus waveguide-coupled microring system with degenerate CW and CCW modes of resonant frequency $\omega_c$, $\hat{a}^{(\dag)}$ [$\hat{b}^{(\dag)}$] are the creation (annihilation) operators of CW [CCW] WGM photons, $\hat{c}_{\mathrm{R}}^{(\dagger)}(x)[\hat{c}^{(\dag)}_{\mathrm{L}}(x)]$ creates (annihilates) a right-moving [left-moving] photon at position $x$ in the bus waveguide with frequency near $\omega_0$, $v_{g}$ is the group velocity of the waveguide mode, $V_{a(b)}=i\sqrt{\kappa_e v_g}$ is the waveguide-microring coupling strength, and $\delta(x)$ is the Dirac delta function. The third line describes a multi-level atom, as illustrated in Fig.~\ref{fig:figLevel}, and its coupling to the CW and CCW WGMs, where $\Omega_{e_l}$ ($\Omega_{g_l}$) is a excited (ground) state energy, $l$ labels the magnetic quantum number, and $\hat{\sigma}_{\pm,l}\equiv\ket{g_l}\bra{e_{l\pm1}}$ is the lowering operator for the corresponding atom-CW (CCW) WGM coupling. 

We consider an atom initially in an arbitrary state $\ket{\Phi(t\rightarrow -\infty)} = \sum_l \alpha_l \ket{g_l}$, and the waveguide is injected with a right propagating photon. In the single excitation limit, evolution of the state should generally take the form
\begin{align}
\ket{\Phi(t)} = &\sum_l \int dx \left(\tilde{\phi}_{\mathrm{R},l}\hat{c}_\mathrm{R}^\dag(x) + \tilde{\phi}_{\mathrm{L},l}\hat{c}_\mathrm{L}^\dag(x) \right) \ket{0}\ket{g_l} \nonumber\\
&+ \sum_l  \tilde{\epsilon}_{e_l}  \ket{0}\ket{e_l} +  \tilde{\epsilon}_{a,g_l} \hat{a}^\dag \ket{0}\ket{g_l} + \tilde{\epsilon}_{b,g_l} \hat{b}^\dag \ket{0}\ket{g_l}\,,
\end{align}
where $\ket{0}$ represents the vacuum state in both the microring and the waveguide, and $(\tilde{\phi}_{\mathrm{R},l},\tilde{\phi}_{\mathrm{L},l}, \tilde{\epsilon}_{e_l}, \tilde{\epsilon}_{a,g_l}, \tilde{\epsilon}_{b,g_l})$ are time-dependent coefficients. We can solve for the state evolution $\ket{\Phi(t)}$  
using the Schr\"odinger equation $i\hbar\frac{\partial}{\partial t}\ket{\Phi(t)} = \hat{H}_{\mathrm{eff}}\ket{\Phi(t)}$ and the effective Hamiltonian Eq.~(\ref{EqSM:sys}). However, we are primarily interested in finding the bus waveguide transmission, and thus the value $\bra{\Phi(t_\mathrm{d})}\hat{c}_\mathrm{R}^\dag(x_\mathrm{d})\hat{c}_\mathrm{R}(x_\mathrm{d})\ket{\Phi(t_\mathrm{d})}$, where the detection point can be taken to infinity $x_\mathrm{d} \rightarrow \infty$ and $t_\mathrm{d} \rightarrow \infty$. We then focus on finding the solution $\ket{\Phi(t)}=\sum_l e^{-i\epsilon_l t}\ket{E_l}$, where the state is written as a superposition of stationary states $\ket{E_l}$, each satisfying $\hat{H}_{\mathrm{eff}}\ket{E_l}=E_l\ket{E_l}$. We find that
\begin{align}
\ket{E_l} = & \int dx \left(\phi_{\mathrm{R},l}\hat{c}_\mathrm{R}^\dag(x) \ket{g_{l}} + \phi_{\mathrm{L},l+2}\hat{c}_\mathrm{L}^\dag(x) \ket{g_{l+2}}\right)\ket{0}  \nonumber\\
&+  \epsilon_{e_{l+1}}  \ket{e_{l+1}} +  \epsilon_{a,g_l} \hat{a}^\dag \ket{0}\ket{g_l} + \epsilon_{b,g_{l+2}} \hat{b}^\dag \ket{0}\ket{g_{l+2}}\,,
\end{align}
and $E_l = \omega + \Omega_{g_l}$, where $\omega=\omega_0 + v_g k$ is the input photon frequency; $\phi_{\mathrm{R},l}(x)=\alpha_l e^{i kx} \left[ \theta(-x) + t_l \theta (x) \right]$, and $\phi_{\mathrm{L},l+2} (x)= \alpha_l r_l e^{-ik_l' x} \theta(-x)$ describe the right- and left-propagating photon wavefunctions in the bus waveguide, respectively, where $k'_l=k+ (\Omega_{g_l}-\Omega_{g_{l+2}})/v_g$, and $\theta(x)$ is the Heaviside step function. Effectively, the system could be described by a collection of $\Lambda$-levels ($\ket{g_l}\leftrightarrow \ket{e_{l+1}}\leftrightarrow \ket{g_{l+2}}$) in the single excitation limit, each giving transmission and reflection coefficients for the right-propagating input photon as
\begin{align}
t_l(\omega,g) = & 1 - \frac{2 \kappa_e }{\tilde{\kappa}} \frac{1+\tilde{C}^{-}_{l}}{1+\tilde{C}^{+}_{l}+\tilde{C}^{-}_{l}} \label{EqSM:tl} \\
r_l(\omega,g) = & \frac{ 2\kappa_e }{\tilde{\kappa}} \frac{\eta^{+}_{l}}{\eta^{-}_{l+2}} \frac{\tilde{C}^{-}_{l}}{1+\tilde{C}^{+}_{l}+\tilde{C}^{-}_{l}}\,,
\end{align}
where $\tilde{C}^{+}_{l}=\frac{4|\eta^{+}_{l}g|^2}{\tilde{\kappa}\tilde{\gamma}_{l}}$ and $\tilde{C}^{-}_{l}=\frac{4|\eta^{-}_{l+2}g|^2}{(\tilde{\kappa}+i\delta_{l}/2)\tilde{\gamma}_{l}}$ are complex cooperativities for $\sigma^{+}$ and $\sigma^{-}$ transitions in the $l$-th $\Lambda$-level, respectively, $\tilde{\kappa} =  \kappa+ 2i\Delta_c$, $\tilde{\gamma}_{l} = \gamma+ 2i(\Omega_{e_{l+1}}-\Omega_{g_{l}} - \omega)$, and $\delta_{l} = \Omega_{g_{l+2}}-\Omega_{g_{l}}$ is the ground state splitting. For $l=F$ and $F-1$ states, $\eta_{l+2}^-=0$, $\tilde{C}^-_l=0$, and $r_l=0$ due to the lack of coupling to the CCW WGM. The total transmission for an atom in state $\ket{\Phi(t\rightarrow -\infty)} = \sum_l \alpha_l \ket{g_l}$ is thus 
\begin{equation}
\mathcal{T}(\omega, g)= \lim_{x_\mathrm{d},t_\mathrm{d}\rightarrow \infty} \left| \bra{\Phi(t_\mathrm{d})}c_\mathrm{R}^\dag(x_\mathrm{d})c_\mathrm{R}(x_\mathrm{d})\ket{\Phi(t_\mathrm{d})}\right|^2 = \sum_l \alpha_l^2|t_l|^2\,,  \label{Eq:SMTavg}  
\end{equation}
and the transmission is the sum of contributions from individual $\Lambda$-levels. 

We note that the coupling to CW and CCW WGMs are asymmetric in the Zeeman manifold. Assuming equal population in the $F=4$ magnetic sub-levels and no Zeeman energy splitting, we have $\braket{\tilde{C}^-}_l = \frac{28}{55}\braket{\tilde{C}^+}_l \approx 0.5 \braket{\tilde{C}^+}_l$, where $\braket{.}_l$ denotes averaging over $l$ states. %We have used this relation for rough estimate of peak transmission $\mathcal{T}/T_0 \approx 2.9$ in Fig.~3 without considering light shifts.

Moreover, in the case of an atom fully polarized in the stretched state, $\ket{F=4, m_F=4}$, transmission spectrum simplifies to 
\begin{align}
\mathcal{T} (\omega,g)  = & |t|^2 = \left|1 - \frac{ 2\kappa_e }{\tilde{\kappa} (1 + \tilde{C})}\right|^2\,, \label{Eq:SMTpol}
\end{align}
and reflectivity $|r|^2=0$. Here $\tilde{C} = 4\frac{g^2}{\tilde{\kappa}\tilde{\gamma}}$, $\tilde{\gamma} = \gamma + 2 i \Delta_a $, and $\Delta_a=  \omega_a - \omega$ is the photon detuning from transition frequency $\omega_a$. In a no-atom case, $\tilde{C}=0$ and the transmission $\mathcal{T} (\omega,0)$ further reduces to that of Eq.~(\ref{Eq:SM:bareT}).

\subsection{Photon statistics of the transmission}

To characterize the quantum character of the microring resonator coupled with guided single atoms, we measure the self-correlation of the transmitted light in a Hanbury-Brown-Twiss (HBT) experiment
\begin{align}
    \label{g2def}
    g^{(2)}(\tau)=\frac{\braket{I_{\mathrm{out}}(t)I_{\mathrm{out}}(t+\tau)}}{\braket{I_{\mathrm{out}}(t)}\braket{I_{\mathrm{out}}(t+\tau)}}=\frac{\braket{\hat{s}_{\mathrm{out}}^\dagger(0){}\hat{s}_{\mathrm{out}}^\dagger(\tau)\hat{s}_{\mathrm{out}}(\tau)\hat{s}_{\mathrm{out}}(0)}}{\braket{\hat{s}_{\mathrm{out}}^\dagger(0)\hat{s}_{\mathrm{out}}(0)}\braket{\hat{s}_{\mathrm{out}}^\dagger(\tau)\hat{s}_{\mathrm{out}}(\tau)}}
\end{align} 
where $\hat{s}_{\mathrm{out}}(t)$ is the output of the microring resonator which is described by the input-output theory $s_{\rm{out}}(t)=s_{\rm{in}}(t)+i\sqrt{\kappa_{\rm{e}}}a(t)\label{Eq: inuputoutput}$. 
For the resonant driving, the analytical result of the time-dependent correlation function $g^{(2)}(\tau)$ of the cavity field $\hat{a}$ has been shown in \cite{StatisticalMethodsInQuantumOptics2} in the bad-cavity limit. We calculate $g^{(2)}(\tau)$ for the transmitted field $\hat{s}_{\mathrm{out}}$ analytically using the quantum regression theorem from the solution of the modified optical Bloch equations. The procedure is same as \cite{StatisticalMethodsInQuantumOptics2} and we take into account the interference of the bare transmission of the resonator and scattering from the atom. The final complete result of is 
\begin{align}
    g^{(2)}(\tau)-1 =& \frac{\kappa_{\rm{e}}^2n_{\rm{sat}}^2}{\left(\braket{\hat{s}_{\mathrm{out}}^\dagger\hat{s}_{\mathrm{out}}}\right)^2}e^{-\frac{3}{4}\gamma'\tau}\left[\Lambda_{+}e^{\delta'\tau}+\Lambda_{-}e^{-\delta'\tau}\right]
\end{align}
where $n_{\rm{sat}}=\frac{\gamma^2}{4g^2}$ is the saturation photon number, $\gamma^{\prime}=(1+C)\gamma=(1+\frac{4g^2}{\kappa\gamma})\gamma$ is the Purcell-enhanced atomic decay rate and $\{\delta^{\prime}, \Lambda_{\pm}\}$ are coefficients
\begin{align}
    \delta^{\prime} =\frac{\gamma'}{4}\sqrt{1-16Y^{'2}}\,&\\
    \Lambda_{+} = \frac{C^2 Y^{\prime 4}}{8 \delta^{\prime}  \left(2 Y^{\prime 2}+1\right)^4} & \left[-C^2 \left[4 \delta^{\prime}  \left(4 \left(Y^{\prime 2}-7\right) Y^{\prime 2}+1\right)+\gamma^{\prime} \left(76 Y^{\prime 4}-52 Y^{\prime 2}+3\right)\right]\right.\nonumber\\
        &+2 (C+1)^2 m_{0}^{2} \left(2 Y^{\prime 2}+1\right)^2 \left(-\gamma^{\prime}-4 \delta^{\prime} +10 \gamma^{\prime} Y^{\prime 2}+8
        \delta^{\prime}  Y^{\prime 2}\right)\nonumber\\
        &\left.-4 (C+1) C m_{0} \left(2 Y^{\prime 2}+1\right) \left(\gamma^{\prime}+4 \delta +8 \gamma^{\prime} Y^{\prime 4}-18 \gamma^{\prime} Y^{\prime 4}-24 \delta^{\prime}  Y^{\prime 2}\right)\right]&\\
    \Lambda_{-} = \frac{C^2 Y^{\prime 4}}{8 \delta^{\prime}  \left(2 Y^{\prime 2}+1\right)^4} & \left[C^2 \left[4 \delta^{\prime}  \left(4 \left(Y^{\prime 2}-7\right) Y^{\prime 2}+1\right)+\gamma^{\prime} \left(76 Y^{\prime 4}-52 Y^{\prime 2}+3\right)\right]\right.\nonumber\\
        &-2 (C+1)^2 m_{0}^{2} \left(2 Y^{\prime 2}+1\right)^2 \left(-\gamma^{\prime}+4 \delta^{\prime} +10 \gamma^{\prime} Y^{\prime 2}-8
        \delta^{\prime}  Y^{\prime 2}\right)\nonumber\\
        &\left.+4 (C+1) C m_{0} \left(2 Y^{\prime 2}+1\right) \left(\gamma^{\prime}-4 \delta +8 \gamma^{\prime} Y^{\prime 4}-18 \gamma^{\prime} Y^{\prime 4}+24 \delta^{\prime}  Y^{\prime 2}\right)\right]
\end{align}
$Y^{\prime}=\frac{Y}{1+C}$ and $Y^2=\frac{2C}{1+m_{0}}|s_{\rm{in}}|^2$ is the dimensionless driving intensity. $m_{0}=\frac{\kappa}{2\kappa_{\rm{e}}}-1$ is related to bare cavity resonant transmission where $m_{0}=\frac{\sqrt{T_0}}{1-\sqrt{T_{0}}}$ for under-coupled condition and $m_{0}=\frac{-\sqrt{T_0}}{1+\sqrt{T_{0}}}$ for over-coupled condition. The output intensity is $\braket{\hat{s}_{\mathrm{out}}^\dagger\hat{s}_{\mathrm{out}}}$ is calculated using the steady-state solution as 
\begin{align}
      \braket{\hat{s}_{\mathrm{out}}^\dagger\hat{s}_{\mathrm{out}}} = \kappa_{\rm{e}} n_{\mathrm{sat}}\frac{2 C Y^{\prime 4}+ \left[m_{0} Y^{\prime} (C+1)  \left(2 Y^{\prime 2}+1\right) +C Y^{\prime}\right]^2}{\left(2 Y^{\prime 2}+1\right)^2}
\end{align}
In the weak driving limit, $Y^{'}\rightarrow 0$, the above expression of $g^{(2)}(\tau)$ can be simplified with $\delta'\rightarrow\frac{\gamma'}{4}$
\begin{align}
        g^{(2)}(\tau)=1-e^{-\frac{3}{4}\gamma'\tau} \frac{C^{2}}{(C+m_{0}C+m_{0})^4} &\left[\left[C^{2}+4m_{0}C(1+C)+2m_{0}^{2}(1+C)^2\right]\cosh(\frac{1}{4}\gamma'\tau)\right.\nonumber\\
        &\left.+\left[3C^{2}+4m_{0}C(1+C)+2m_{0}^{2}(1+C)^2\right]\sinh(\frac{1}{4}\gamma'\tau)\right]
\end{align}
With perfect critical coupling condition $m_{0}=0$, expression of $g^{(2)}(\tau)$can be further simplified to 
\begin{align}
    g^{(2)}(\tau)&=1-e^{-\frac{3}{4}\gamma'\tau}\left[\cosh(\frac{1}{4}\gamma'\tau)+3\sinh(\frac{1}{4}\gamma'\tau)\right]\nonumber\\
    &=(1-e^{-\frac{1}{2}\gamma'\tau})^2\,,
\end{align}
which is identical with the formula in Ref.~\cite{2008Science_Kimble}.

\section{Fitting procedure for guided atoms}
\setcounter{equation}{28}
We use Eqs.~(\ref{Eq:SMTavg}) and (\ref{Eq:SMTpol}) to fit the transmission spectrum of unpolarized guided atoms and spin-polarized guided atoms, respectively. For guided atoms, we calculate a time-averaged transmission spectrum since atom-WGM photon interaction occurs during an atom in transit through the near field, and the coupling strength $g$ 
and the atomic energy levels ($\Omega_{e_l}, \Omega_{g_l}$) are position dependent. We incorporate trajectory and light shift calculations to obtain an ensemble of time-dependent ($g(t), \Omega_{e_l}(t), \Omega_{g_l}(t)$). This allows us to calculate the averaged transmission with the occurrence of $\mathcal{N}$ atom-transits during the 1~ms probe window. Using Eq.~(\ref{Eq:SMTavg}) and (\ref{EqSM:tl}), we calculate 
\begin{equation}
    \frac{T(\omega)}{T_0}= 1+ \mathcal{N}\left[\frac{\braket{\int dt\,\mathcal{T}(\omega, g(t))}}{\int dt\,\mathcal{T}(\omega, 0)} - 1\right]\,, \label{EqSM:fitmodel}
\end{equation}
where $\braket{...}$ denotes ensemble averaging over all simulated trajectories. For unpolarized guided atoms, we have assumed equal population in all $m_F$ states. This model assumes $\mathcal{N}$ as the only fit parameter, and captures the spectral line shape very well as shown in Fig.~3.

\section{Data analysis procedure for Fig. 4}
\setcounter{equation}{29}
We describe our data analysis procedure for Fig. 4. Our procedure is similar to that found in Ref.~\cite{2008Science_Kimble}. Figure 4(a) is calculated using the outputs ($C_{1}(t)$ and $C_{2}(t)$) of two SPCMs directly without any selection. Increased count rate due to atom arrival within time $4~\mathrm{ms}\leq \Delta t\leq 8~\mathrm{ms}$ is clearly visible. 

\begin{figure}[ht!]
\centering
\includegraphics[width=1.0\columnwidth]{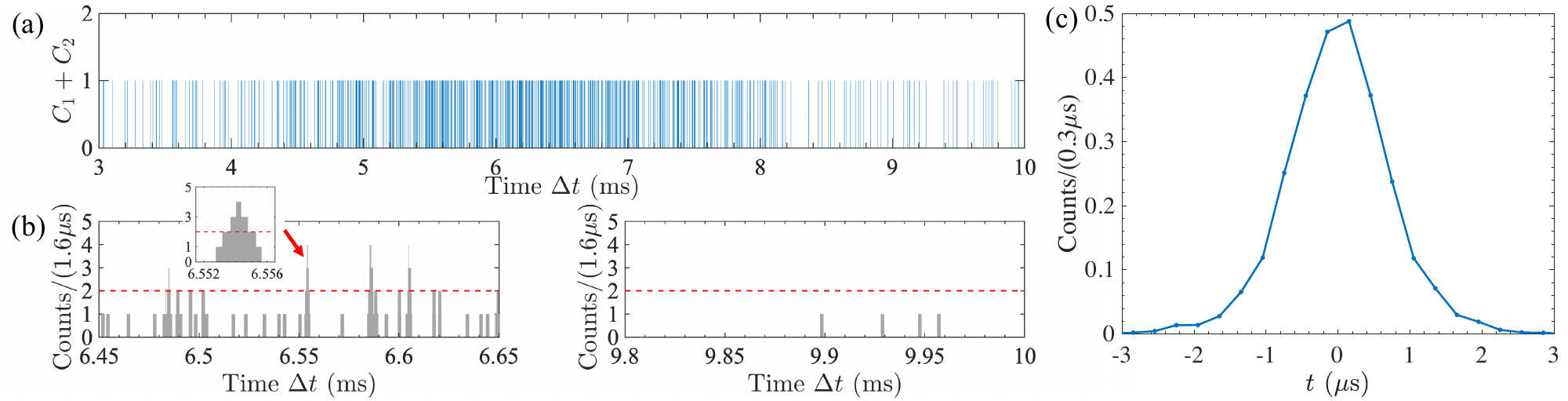}
\caption{ (a) Recorded time-stamped photon counts with different guiding time $\Delta t$ from the combined outputs ($C_1$ and $C_2$) of two SPCMs with time bins of $0.8~$ns. (b) Photon count rate (gray) within a $1.6~\mu s$ running window using the data around $6~$ms (Left) and $10~$ms (Right) in (a). Atom-transit events are selected with a threshold of 2 counts per $1.6~\mu s$(red dashed line). Inset shows an identified atom-transit event indicated by the red arrow. (c) Averaged photon count ($0.3~\mu s$ time bin) from the selected atom-transit events. %Time origin $t=0$ of each atom-transit is determined by averaging the time weighted by the selected actual counts. 
}\label{fig:figcount}
\end{figure}

To extract the correlation of transmitted photons with an atom coupled to the WGM of the microring resonator, we monitor the combined counts $\sum_{i}\left[C_{1}(t_{i})+C_{2}(t_{i})\right]$ within a $1.6~\mu$s sliding window and apply a selection criterion to identify single atom transits. Figure \ref{fig:figcount}(b) compares counts measured at guiding time $\Delta t\approx 6~$ms and $10~$ms, respectively, illustrating the difference with and without guided atoms. We identify traces with peak counts $\geq 2$ as clear atom-transit events and select them for data analyses. We determine the center of each transit event by averaging time weighted by the counts. %We then overlap the center of each identified transit events at $t=0$. 
Averaged photon count rate around the center of each transit event (now shifted to $t=0$) is shown in Fig. \ref{fig:figcount}(c). Using the photon count traces $C_{1}$ and $C_2$ around the center of each selected transit event ($|t|\leq 1~\mu$s), we calculate the correlation of transmitted photons $g^{(2)}(\tau)$ in the presence of atoms. The results are shown in Fig. 4(b-c).
%single atom-transit events with increased photon count rate are clearly visible for . After selection, only segments of $C_{1,2}(t)$ which gives high photon count rates are remained and the time origin $t=0$ of each segments (the center of each atom-transit event) is determined . 
%With the identified atom-transit events, 
\bibliography{apssamp}